# Editorial: Spatial Heterogeneity, Scale, Data Character, and Sustainable Transport in the Big Data Era

Bin Jiang
Faculty of Engineering and Sustainable Development, Division of GIScience
University of Gävle, SE-801 76 Gävle, Sweden
Email: bin.jiang@hig.se

*(Draft: April 2018, Revision: April, May, June 2018)*

In light of the emergence of big data, I have advocated and argued for a paradigm shift from Tobler's law to scaling law, from Euclidean geometry to fractal geometry, from Gaussian statistics to Paretian statistics, and – more importantly – from Descartes' mechanistic thinking to Alexander's organic thinking. Fractal geometry falls under the third definition of fractal – that is, *a set or pattern is fractal if the scaling of far more small things than large ones recurs multiple times* (Jiang and Yin 2014) – rather than under the second definition of fractal, which requires a power law between scales and details (Mandelbrot 1982). The new fractal geometry is more towards living geometry that *"follows the rules, constraints, and contingent conditions that are, inevitably, encountered in the real world"* (Alexander et al. 2012, p. 395), not only for understanding complexity, but also for creating complex or living structure (Alexander 2002–2005). This editorial attempts to clarify why the paradigm shift is essential and to elaborate on several concepts, including spatial heterogeneity (scaling law), scale (or the fourth meaning of scale), data character (in contrast to data quality), and sustainable transport in the big data era.

Table 1: Comparison of the scaling law versus Tobler's law
(Note: These two laws complement each other and recur at different levels of scale in geographic space or the Earth's surface.)

| Scaling law | Tobler's law |
|---|---|
| far more small things than large ones | more or less similar things |
| across all scales | available on one scale |
| without an average scale (Pareto distribution) | with an average scale (Gauss distribution) |
| long tailed | short tailed |
| interdependence or spatial heterogeneity | spatial dependence or homogeneity |
| disproportion (80/20) | proportion (50/50) |
| complexity | simplicity |
| non-equilibrium | equilibrium |

Current geographic information systems (GIS), which were first conceived and developed in the 1970s, are still largely based on the legacy of conventional cartography (Goodchild 2018). Although computer technology has advanced dramatically since then, the legacy or the fundamental ways of thinking remains unchanged. For example, GIS representations of raster and vector and even so-called object-oriented representation are still constrained among geometric primitives such as pixels, points, lines, and polygons (Longley et al. 2015). These geometry-oriented representations help us to see things that are more or less similar, characterized by Tobler's law (1970), or commonly known as the first law of geography. For example, the price of your house may be similar to those of your neighbors, but there are far more low house prices than high ones. This notion of far more lows than highs – or far more smalls than larges in general – is what underlies scaling law for characterizing spatial heterogeneity. The concept of spatial heterogeneity, as conceived in current geography literature, is mistaken because it does not recognize the fact of far more smalls than larges (Jiang 2015b). This notion of far more smalls than larges adds a fourth meaning of scale; that is, a series of scales ranging from the smallest to the largest form the scaling hierarchy (Jiang and Brandt 2016). The scaling hierarchy can be further rephrased as: numerous smallest, and a very few largest, and some in between the smallest and the largest. In order to see far more smalls than large one, we must adopt a topological perspective on meaningful geographic features such as streets and cities instead of the geometric primitives. In other words, we must shift our eyes from geometric details to overall data character (which will be further elaborated on below). Tobler's law depicts a fact as a local scale. However, geographic space is governed by not only Tobler's law but also by scaling law. These two laws are complementary to each other (Table 1). Calling for a shift from Tobler's law to scaling law is not to abandon Tobler's law, but to shift from thinking that is dominated by Tobler's law to thinking dominated by scaling law because scaling law is universal and global.

Benoit Mandelbrot, the father of fractal geometry, remarked that unlike things we see in nature, Euclidean shapes are cold and dry. Christopher Alexander, the father of living geometry, referred to structures with a higher degree of wholeness as living structures. Wholeness is defined mathematically as a recursive structure, and it exists in space and matter physically and reflects in our minds and cognition psychologically (Alexander 2002–2005, Jiang 2016). A cold and dry structure versus a living one vividly describes the difference between Euclidean shapes and fractal or living structures. A shift from Euclidean geometry to fractal or living geometry is not to abandon Euclidean geometry, but Euclidean geometric thinking. Euclidean geometry is essential for fractal geometry, since one must first measure it in order to see whether there are far more smalls than larges. However, Euclidean geometric thinking differs fundamentally from that of fractal geometry. For example, Euclidean geometric thinking tends to see things individually (rather than holistically) and non-recursively (rather than recursively); refer to Jiang and Brandt (2016) for a more detailed comparison about these two geometric ways of thinking. The shift from Euclidean to fractal or living geometry implies that fractal or living geometry is to be the dominant way of thinking. To present a practical example, a cartographic curve is commonly seen, under the Euclidean geometric thinking, as a set of more or less similar line segments; however, it should more correctly be viewed, under the fractal or living geometric thinking, as a set of far more small bends than large ones, and importantly small bends are recursively embedded in the large ones.

According to Tobler's law, the price of your house is similar to those of your neighbors. In other words, averaging your neighbors' housing prices would lead to your housing price. In this case, the average makes a good sense for predicting your housing price, as all neighboring housing prices can be characterized by a well-defined mean. In order for the prediction to make sense, there is a condition to meet; namely, the neighboring housing prices are more or less, or with a well-defined mean. This is indeed true for housing prices on a local scale. This condition is violated on a global scale, since there are far more low prices than high ones. In this case, the average does not make good sense; it lacks an average or scale-free or scaling. Things with more or less similar sizes can be well modeled by Gaussian statistics, whereas things with far more smalls than larges should be well characterized by Paretian statistics. In this regard, we have developed a new classification scheme for data with far more smalls than larges. This classification scheme is called head/tail breaks (Jiang 2013), which recursively breaks data into the head (for data values greater than an average) and the tail (for data values less than an average) until the condition of a small head and a long tail is violated. A head/tail breaks-induced index called the ht-index (Jiang and Yin 2014) can be used to characterize the notion of far more smalls than larges or underlying scaling hierarchy. As mentioned above, the new definition of fractal is based on the notion of far more smalls than larges.

Under Euclidean geometric thinking that focuses on geometric details, data quality or uncertainty is one of the priority issues in GIS. This kind of thinking is evident in many scientific papers and talks about, for example, data quality of OpenStreetMap (OSM). I believe that the GIS field has over-emphasized the data quality issue, so I would like to add a different view, arguing that data quality is less important than data character. By data character, I mean some overall character of geospatial data, or the wholeness, or living structure as briefly mentioned above. To illustrate, this link (https://twitter.com/binjiangxp/status/985322539625967618) shows a cartoon and a photo of Kim Jong-Un. The photo on the right has the highest geometric details, and the cartoon on the left has the lowest geometric details. However, the cartoon on the left captures the highest character or personality.

This link (https://twitter.com/binjiangxp/status/985322961342263296) further illustrates the living structure of the street network of a small neighborhood. The street network on the right has the highest data quality, while the graph on the left captures the highest data character – far more less-connected streets than well-connected ones. What I want to argue is that if the street network on the right suffers from some errors, this would not have much effect on the data character on the left. It is in this sense that quality is not super-important compared to data character. If data quality or geometric details were compared to trees, then data character would be the forest. To present a specific example, Jiang and Liu (2012) adopted a topological perspective and examined the scaling of geographic space based on OSM data of three European countries: France, Germany, and the UK. There is little doubt that there were numerous errors in the OSM data, but they have little effect on the finding – the scaling or living structure of geographic space in which there are far more small things than large ones.

The legacy of GIS has been driven substantially by the mechanistic thinking of over past 300 years of science (Descartes 1637, 1954). Everything we have achieved in science and technology benefits greatly from mechanistic thinking, but it is limited in terms of how to make a better built environment (Alexander 2002–2005). This mechanistic thinking is reflected in the GIS representations of raster and vector, and in the box counting for calculating fractal dimensions. It is also reflected in top-down imposed geographic units such as census tracts; these are clearly very useful for administration and management, but of little use for scientific purposes. Space is neither lifeless nor neutral, but has a capacity to be more living or less living (Alexander 2002–2005). In other words, space is a living structure with a high degree of wholeness. A country is a living structure that consists of far more small cities than large ones. Figure 1 shows all the natural cities of Austria, derived from street nodes of the country's OSM data. Seen from the figure, all cities have very natural boundaries, and they are quite coherent, with a topographic surface that reveals the underlying living structure of far more small cities than large ones. Natural cities are objectively defined cities, from a massive number of geographic locations, such as social media locations (Jiang and Miao 2015); please refer to the Appendix of the paper for details on the derivation of natural cities.

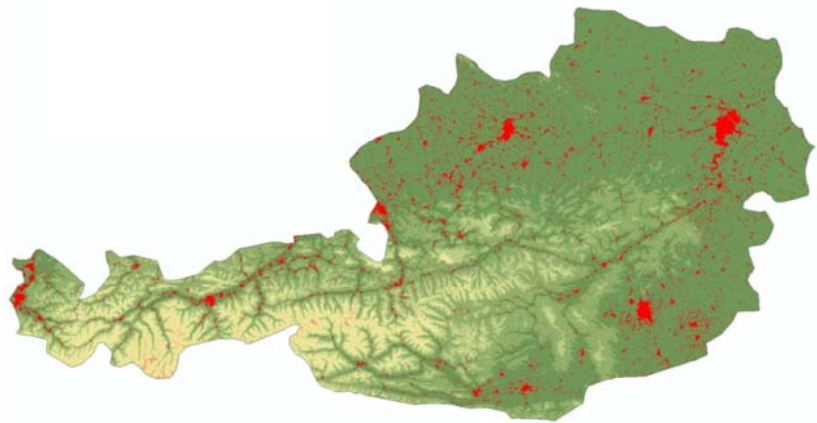

Figure 1: (Color online) Living structure of all natural cities derived from Austria's OSM data
(Note: This topographic rendering is based on head/tail breaks (Jiang 2013, 2015a). Unlike traditional renderings, this head/tail breaks induced rendering clearly shows the living structure of far more low elevations than high ones.)

Geographic space is a living structure, not just at the country scale, but also at the city scale. Figure 2 shows natural streets of Vienna and Linz, which demonstrate striking living structures with far more less-connected streets than well-connected ones. The natural streets are able to capture the underlying scaling or living structure of far more the less-connected than the well-connected, and are therefore able to predict up to 80 percent of traffic flow. In other words, traffic flow is mainly shaped by the living structure and has little to do with human travel behavior. In this circumstance, human beings can be thought of as atoms or molecules that interact with each other, and with the natural streets to shape the traffic flow. Traffic is not a phenomenon, but an outcome of the living structure. This is in line with the famous statement by Winston Churchill – *We shape our buildings and afterwards our* *buildings shape us*. With respect to sustainable transport, we can paraphrase Churchill: We shape our transport system, and it will shape us, so make sure we shape it well so that we will be well-shaped too. To be more specific, we shape our transport system as a living structure and it then shapes our sustainable mobility.

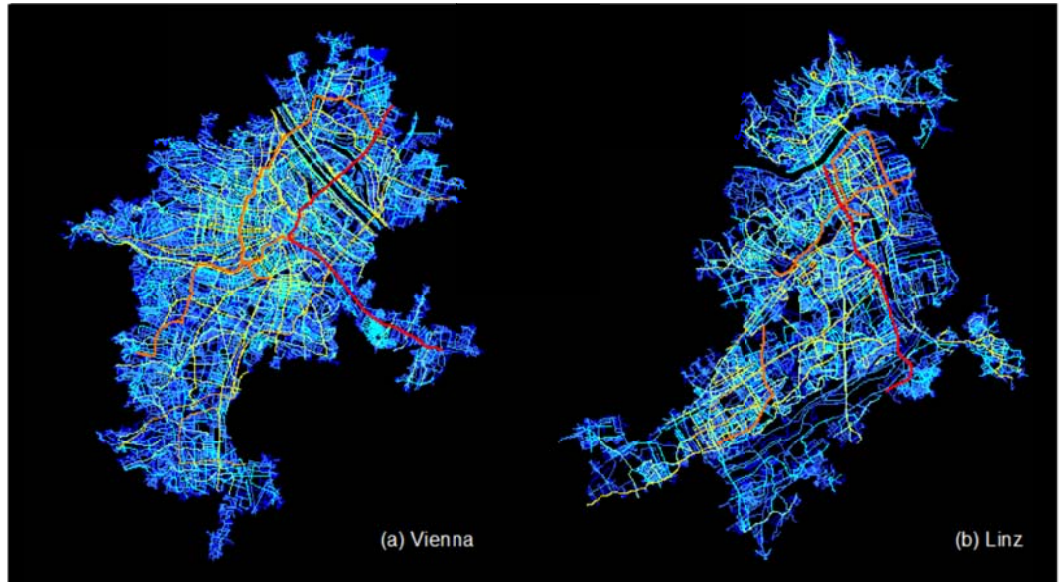

Figure 2: (Color online) Living structure of natural streets of (a) Vienna and (b) Linz
(Note: The living structure shows far more less-connected streets – indicated by many cold colors than well-connected ones – indicated by a very few warm colors.)

This social physics perspective offers new insights into traffic flow. To this point, I would like to end this editorial with the following excerpt (Buchanan 2007):

*"There's an old way of thinking that says the social world is complicated because people are complicated … We should think of people as if they were atoms or molecules following fairly simple rules and try to learn the patterns to which those rules lead … Seemingly complicated social happenings may often have quite simple origins … It's often not the parts but the pattern that is most important, and so it is with people."*

**Acknowledgement**
This editorial was substantially inspired by my recent panel presentation *"On Spatiotemporal Thinking: Spatial heterogeneity, scale, and data character"*, presented at the panel session entitled "Spatiotemporal Study: Achievements, Gaps, and Future" with the AAG 2018 Annual Meeting, New Orleans, April 10–15, 2018, and my keynote *"A Geospatial Perspective on Sustainable Urban Mobility in the Era of BIG Data"*, presented at CSUM 2018: Conference on Sustainable Urban Mobility, May 24–25, Skiathos Island, Greece.